# Thermodynamic and electron-transport properties of $Ca_3Ru_2O_7$

## from first-principles phonon calculations and Boltzmann transport theory

Yi Wang[1*], Yihuang Xiong[1†], Tiannan Yang[1], Yakun Yuan[2], Shunli Shang[1], Zi-Kui Liu[1],

Venkatraman Gopalan[1], Ismaila Dabo[1], and Long-Qing Chen[1]

[1]Department of Materials Science and Engineering, The Pennsylvania State University,

University Park, PA 16802, USA

[2]School of Department of Mechanical Engineering and Zhangjiang Institute for Advanced Study,

Shanghai Jiao Tong University, Shanghai, China

This work demonstrates a first-principles-based approach to obtaining finite temperature thermal and electronic transport properties which can be employed to model and understand mesoscale structural evolution during electronic, magnetic, and structural phase transitions. A computationally tractable model was introduced to estimate the temperature dependence of the electron relaxation time. The model is applied to $Ca_3Ru_2O_7$ with a focus on understanding its electrical resistivity across the electronic phase transition at 48 K. A quasiharmonic phonon approach to the lattice vibrations was employed to account for thermal expansion while the Boltzmann transport theory including spin-orbit coupling was used to calculate the electron-transport properties, including the temperature dependence of electrical conductivity.

---

[*] yuw3@psu.edu;
[†] yyx5048@psu.edu

# 1    Introduction

Due to their unconventional magnetic and electronic properties, Ruddlesden-Popper (R-P) ruthenates $(Sr,Ca)_{n+1}Ru_nO_{3n+1}$ are attracting increasing interest in the field of solid-state physics and materials science [1]. Notably, $Ca_3Ru_2O_7$ is one of the few known polar metals (which are able to retain a spontaneous electric polarization in the metallic state) [2]. In its $Bb2_1m$ crystalline form [3], $Ca_3Ru_2O_7$ exhibits a rich variety of physical phenomena, including temperature-dependent band dispersion [4–6], pressure-induced magnetic phase transition [3], colossal magnetoresistance [7], strong correlation, and pronounced spin-orbit coupling, making it a prototypical system to study the effects of temperature on the electronic, magnetic, and transport properties of polar metals. Cooled down below its Néel temperature of 56 K, $Ca_3Ru_2O_7$ becomes antiferromagnetic with spins aligned along its $a$-axis (AFM-$a$). When further cooled down to 48 K, it undergoes a second magnetic phase transition, where spins reorient along the $b$-axis (AFM-$b$); this transition is accompanied by an isostructural phase transformation (corresponding to a contraction of the unit cell along its $c$-axis) and by a sudden change in resistivity of semimetallic character (often interpreted as arising from the the opening of a pseudo-gap) [5,6,8]. Below 30 K, $Ca_3Ru_2O_7$ undergoes another phase transition whereby it recovers its metallic temperature-dependent resistivity.

While first-principles calculations based on density functional theory (DFT) [9,10] have demonstrated their accuracy in predicting lattice vibrations, electron excitations, and configuration effects [11–14], it is still an ongoing challenge to evaluate the transport properties of materials. For instance, the calculations of electrical conductivities typically rely on the Boltzmann transport theory [15,16] based on estimated electron relaxation times. While these relaxation times can be calculated using the Bardeen-Shockley deformation-potential theory [17,18] under the effective mass approximation together with phenomenological parameters [19,20], the majority of DFT-based calculations [15,21,22] treat them as a fitting parameter whose values are found to be on the order of $10^{-14}$ s [24].

This work reports the thermal and electrical properties of $Ca_3Ru_2O_7$ from first-principles calculations based on density-functional theory. In this work, first-principles quasiharmonic phonon calculations are carried out to understand the thermodynamic and electrical properties of $Ca_3Ru_2O_7$. A tractable model is proposed to estimate the temperature dependence of the electron relaxation time by correlating electron-relaxation times to the specific heat per mobile charge, as initially suggested by the previous work [23,24].

## 2    Quasiharmonic approach

The Helmholtz energy of a crystal [11,25] can be expressed as

$$F(V, T) = E_c(V) + F_{vib}(V, T) + F_{el}(V, T)$$    **Eq. 1**

where $V$ is the volume, $T$ is the temperature, $E_c$ is the equilibrium total energy at $T = 0$ K, $F_{vib}$ is the vibrational contribution, and $F_{el}$ is the thermal electronic contribution to the free energy. The Gibbs energy can be calculated as $G = F + PV$, where $P$ is the pressure. Here, we will limit the analysis to systems under zero external pressure ($G = F$). $E_c$ can be obtained from conventional DFT geometry optimization; $F_{vib}$ can be calculated from the DFT phonon spectrum [11] or using the Debye model [26]; and $F_{el}$ can be calculated by the Mermin's finite-temperature density functional theory [25,27] as

$$F_{el}(V, T) = E_{el}(V, T) - TS_{el}(V, T)$$    **Eq. 2**

where $S_{el}$ is the bare electronic entropy [27,28]:

$$S_{\text{el}}(V,T) = -k_{\text{B}} \int D(\varepsilon)[f(\varepsilon, V, T) \ln(f(\varepsilon, V, T))$$

$$+ (1 - f(\varepsilon, V, T)) \ln(1 - f(\varepsilon, V, T))]d\varepsilon$$

**Eq. 3**

where $D(\varepsilon)$ is the electronic density-of-states given by

$$D(\varepsilon) = \int \sum_i \delta(\varepsilon - \varepsilon_i(\boldsymbol{k})) \frac{d\boldsymbol{k}}{8\pi^3}$$

**Eq. 4**

where the index $i$ runs over the electronic bands, and $\boldsymbol{k}$ represents the Bloch wavevector in reciprocal space. $f$ in Eq. 5 is the familiar Fermi distribution,

$$f = \frac{1}{\exp\left(\dfrac{\varepsilon - \mu(T)}{k_B T}\right) + 1}$$

**Eq. 5**

where $\mu(T)$ represents the chemical potential of electrons at finite temperature, i.e., the Fermi level.

The thermal electron energy takes the form

$$E_{\text{el}}(V,T) = \int D(\varepsilon) f(\varepsilon, V, T) \varepsilon d\varepsilon - \int^{\varepsilon_F} D(\varepsilon) \varepsilon d\varepsilon$$

**Eq. 6**

where $\varepsilon_F$ is the Fermi energy at 0 K.

## 3    Boltzmann transport theory

The electrical conductivity in the Boltzmann transport theory is written as

$$\boldsymbol{\sigma} = \frac{e^2}{V k_{\text{B}} T} \int_{-\infty}^{\infty} f(1 - f)\,\Xi(\varepsilon) d\varepsilon$$

**Eq. 7**

where $e$ is the elementary charge, and $\Xi(\varepsilon)$ is the so-called the transport function [15,16]. $\Xi(\varepsilon)$ is a tensor with components

$$\Xi^{\alpha\beta}(\varepsilon) = \int \sum_i \tau_{i,\boldsymbol{k}} v_i^\alpha(\boldsymbol{k}) v_i^\beta(\boldsymbol{k})\, \delta(\varepsilon - \varepsilon_i(\boldsymbol{k})) \frac{d\boldsymbol{k}}{8\pi^3} \qquad \textbf{Eq. 8}$$

where $\alpha$ and $\beta$ are the indices labeling the cartesian axis, $\tau_{i,\boldsymbol{k}}$ is electron relaxation time, and the electron group velocity $v_i^\alpha$ is the gradient of electron band energy with respect to $\boldsymbol{k}$, namely

$$v_i^\alpha(\boldsymbol{k}) = \frac{1}{\hbar} \frac{\partial \varepsilon_i(\boldsymbol{k})}{\partial k^\alpha} \qquad \textbf{Eq. 9}$$

## 4    Estimating electron relaxation time

### 4.1    Consideration based on Heisenberg uncertainty principle

In this section, we discuss the estimation of the electron relaxation time based on DFT calculations. The idea stems from a consideration based on the Heisenberg uncertainty principle in addition to the observation that the thermal energy per mobile charge carrier is on the scale of $k_B T$. One can then guess that the electron relaxation time could be roughly on the scale of $\hbar/(2k_B T)$. As a matter of fact, at 300 K, this results in a value of $1.27 \times 10^{-14}\ s$ which is very close to the commonly assumed value of $1.0 \times 10^{-14}\ s$ for the electron relaxation time in literature [15,16,21].

We will follow the constant electron relaxation time approximation, i.e., treat $\tau_{i,\boldsymbol{k}} = \tau$ in **Eq. 8**. The inspiration is from the Heisenberg uncertainty principle which imposes the lower limit for the product between the measurable uncertainty of energy and the measurable uncertainty of time by

$$\langle \Delta \varepsilon \rangle \cdot \langle \Delta t \rangle \geq \frac{\hbar}{2} \qquad \textbf{Eq. 10}$$

If we can approximate the electron relaxation time $\tau \propto x\langle\Delta t\rangle$, we have

$$\langle\Delta\varepsilon\rangle \cdot \tau = x\frac{\hbar}{2} \qquad\qquad \textbf{Eq. 11}$$

where $x$ is a material specific constant. To get $\langle\Delta\varepsilon\rangle$, one is inspired by the heat capacity formulation in terms of the statistical fluctuation [29] of the thermal energy from which we relate $\langle\Delta\varepsilon\rangle$ to the heat capacity of electrons by

$$\frac{c_{el}}{n} = \frac{\langle(\varepsilon - \langle\varepsilon\rangle)^2\rangle}{k_B T^2} \cong \frac{\langle\Delta\varepsilon\rangle^2}{k_B T^2} \qquad\qquad \textbf{Eq. 12}$$

where $c_{el}$ is the electronic contribution to the specific heat, and

$$n = \int_{-\infty}^{\infty} (1-f)f\, D(\varepsilon)d\varepsilon \qquad\qquad \textbf{Eq. 13}$$

$n$ can be considered as the number of the mobile charge carriers, or the number of active electronic thermal carriers. Eq. 13 shows that the electronic states near the Fermi level [$\mu(T)$ in Eq. 5] [27,30] contributes the most to the electric or thermal conduction as it is dictated by the factor of $(1-f)f$ which mimics an interaction between electron and hole states through $f$ and $(1-f)$, respectively. In other words, the electron system can be viewed as a system made up of mobile charge carriers which makes the main contributions to the electronic heat conductivity, electronic heat capacity, and electric conductivity.

$\langle\varepsilon\rangle$ in **Eq. 12** is the average band energy per mobile charge carrier defined as

$$\langle\varepsilon\rangle = \frac{1}{n}\int_{-\infty}^{\infty} \varepsilon(1-f)f D(\varepsilon)d\varepsilon \qquad\qquad \textbf{Eq. 14}$$

$c_{el}$ in **Eq. 12** can be calculated by

$$c_{el} = \frac{1}{k_B T^2} \int_{-\infty}^{\infty} (\varepsilon - \langle \varepsilon \rangle)^2 (1-f) f D(\varepsilon) d\varepsilon \qquad \textbf{Eq. 15}$$

Finally, substituting **Eq. 12** into **Eq. 11**, one gets

$$\tau = x \frac{\hbar}{2T} \sqrt{\frac{n}{k_B c_{el}}} \qquad \textbf{Eq. 16}$$

Note that $n$ and $c_{el}$ can be calculated using **Eq. 13** and **Eq. 15**, respectively. Treating $x$ in **Eq. 11** as a fitting parameter, we found $x = 0.5$ is a good choice for $Ca_3Ru_2O_7$ in its two low temperature phases.

Theoretically, the electron relaxation time was mostly analyzed in terms of the rates of impurity, acoustic phonon, and polar phonon scattering [31,32] as well as electron-electron scattering [5]. The resulting electron relaxation time ($\tau$) in **Eq. 16** could be considered as an effective estimate incorporating all these scatterings in an average way.

## 4.2 Supportive evidence from previous literaure

In a separate work [24], we proved that $c_{el}/n$ in **Eq. 12** is related to the Lorenz number [33,34] by a factor of $k_B/e^2$. Considering the fact that the Lorenz number was weakly temperature dependent which was especially true for metallic materials [35–38], it was observed from **Eq. 16** that the relaxation time by the present work was virtually inversely proportional to the temperature. This temperature proportionality is the same with the recent works, such as refs. [19,39] report $\tau = C n^{-1/3}/T$ where $n$ is the doping level, and $C$ is a fitting parameter, Wilson and Block's result for metals [40,41], the Umklapp process reported in ref. [42], and Ziman's results [43,44].

# 5    Computational details

## 5.1    Electronic-structure calculations

DFT calculations are performed using the Vienna Ab-initio Simulation Package (VASP) with considering spin-orbit interactions. The projected augmented wave method [45,46]. LDA [47] (local density approximation) functional is utilized to assess the electron and phonon properties. To account for the strong correlation among the $d$ electrons in Ru, the on-site Coulomb repulsion of 1.2 eV is applied on the $4d$ orbitals using the Dudarev's approach [48]. The initial lattice parameters are taken from experimental measurements [1] at 8 K and 50 K, respectively, which correspond to the AFM-$b$ and AFM-$a$ magnetic ordering. The optimization of the atomic positions is carried out with a plane-wave cutoff of 650 eV, and the Brillouin zone is sampled using Gaussian smearing with a 20 meV width on a 5×5×3 Γ-centered $k$-mesh. The energy and forces are converged to be within $10^{-8}$ eV and 0.1 meV/Å. After the self-consistent calculations, non-self-consistent calculations are performed using denser $k$-mesh of 10x10x6 for more accurate electronic energy eigenvalues to calculate the transport properties of electrons based on the Boltzmann transport theory [15,16].

## 5.2    Computational implementation

The workflow for the computerization is given in **Figure 1**. To implement the formulism, we modified the BoltzTrap2 code [49] by adding the functions to calculate the electron heat capacity and effective charge carrier density as described in **Eq. 12** and **Eq. 13**. To make sure the computational accuracy at low temperature region, the mesh for the one-electron energy was modified from uniformly sampling to a self-adapted sampling with denser mesh (1000 time) near the Fermi energy by Gaussian distribution. The procedure for calculating the chemical potential of electron was also revised by implementing the Brent's method [50] to improve computational efficiency.

The thermodynamic calculations are performed using the DFFTK package [51] which has been released to the public under the MIT software license. In addition to the routine calculations of thermodynamic properties via the quasiharmonic approach (QHA) [11,52], it has been implemented in the DFTTK that any properties, as long as they depend on volume or stain, can be calculated under a quasi-static approach via the predicted property-volume/strain relationship from the QHA [25,52]. Therefore, the effects of thermal expansion have been considered for calculating both the electron relaxation time and the electrical conductivity.

## 6    Results and discussions

### 6.1    Heat capacity and Debye temperature

The calculated heat capacities for the AFM-$a$ and AFM-$b$ phases of $Ca_3Ru_2O_7$ are compared with a collection of experimental data [1,53,54] in Figure 2. It shows excellent agreements between the calculations and experiments except for the experimental spike around 48 K. A heat capacity spike in the vicinity of a phase transition temperature is typical for structural phase transitions. The thermal electronic contribution in **Eq. 15** is separated from the lattice contribution as

$$C_{p,lat+el} = c_{el} + C_{p,lat} \qquad\qquad \textbf{Eq. 17}$$

Figure 2 shows that the electronic contributions are small.

Next, we investigate the behaviors of the heat capacity at low temperature region as routinely performed [1,54] via the form of $C/T$ vs $T^2$, namely,

$$C_{p,lat+el}/T = \gamma + \beta T^2 \qquad \qquad \textbf{Eq. 18}$$

where $\gamma$ is the so-called electronic heat capacity coefficient [55], and based on the value of $\beta$ one can calculate the Debye temperature. At the low temperature limit, the calculated $\gamma$'s by the present work are 0.23 mJ/mol-atom and 0.90 mJ/mol-atom, for the AFM-*b* and the AFM-*a* phases, respectively. In particular, the value of 0.23 mJ/mol-atom for the AFM-*b* phase agrees excellently with the calorimetric result reported by Ke et al. [53] and is close to the value of 0.28 mJ/mol-atom reported by Yoshida et al. [1], whereas it is one magnitude smaller than the reported value of 3.6 mJ/mol-atom by McCall et al. [54] and Gao et al. [56].

Approaching to the 0 K limit, we get the Debye temperatures of 492.4 K and 476.4 K, for the AFM-*b* and AFM-*a* phases, respectively. In comparison, the reported Debye temperature by McCall et al. [54] was 480 K based on fitting their measurements. Away from the low temperature region, one can get the Debye temperature by fitting the calculated constant heat capacity from the phonon approach utilizing the Debye formula for the heat capacity [26,57]. **Figure 2** shows that the Debye temperatures are moderately temperature dependent.

## 6.2    Calculated physical quantities from the electron density of states

Major thermal properties of electrons can be calculated from the electron density of states [12]. The electron density of states (DOS) calculated for the AFM-*b* and AFM-*a* phases are illustrated in **Figure 3**a. At the Fermi energy, the DOS for the AFM-*b* is roughly half of that of the AFM-*a* phase. This ratio is quite similar to the measured ratio of the electrical conductivity [5] of the AFM-*b* to the AFM-*a* phases. It is observed that the opposite behaviors [5] on the locations of Fermi energies for the two phases, i.e., a dip structure for the AFM-*b* phase vs a peak structure for the AFM-*a* phase at the Fermi energy. This observation could correspond to the experimental suggestion of the appearance of an insulating-like pseudo-gap [5].

Hereby want to reiterate the importance of concept of "mobile charge carriers" as given in **Eq. 13** which was introduced in a previous work [23]. On one hand, it showed that only the electronic states with energies around the Fermi level can contribute to the thermal properties, by a factor of $f(1-f)$ to the electron density of states as seen from **Eq. 13**, **Eq. 14**, and **Eq. 15**. As a matter of fact, $f(1-f)$ behaves quite like a Dirac delta function except a normalization factor when approaching to low temperature. For the two AFM phases of $Ca_3Ru_2O_7$, the calculated mobile charge carriers are illustrated in **Figure 3**b which shows that the calculated densities of mobile charge carriers for the two phases are nearly linear temperature dependent, typical for metallic materials.

The most important quantity came into the expression for electron relaxation time in **Eq. 16** is the electronic heat capacity per effective mobile charge carriers, namely $c_{el}/n$ in **Eq. 12**. The calculated $c_{el}/n$'s for the two phases of $Ca_3Ru_2O_7$ are plotted in **Figure 3**c. It shows that the values of $c_{el}/n$'s are roughly constants. This can be understood in terms of the Lorenz number which is a factor of $k_B/e^2$ to $c_{el}/n$ as we proved in a separate work [24]. Last, plotted in **Figure 3**d is the estimated electron relaxation time based on **Eq. 16** using $x$=0.5 which is found to be a good fit to match the electrical resistivity measured by Yuan et al. [5].

### 6.3    Calculated physical quantities from the transport electron density of states

The transport electron density of states is a fundamental quantity to calculate almost all key kinetic properties of electrons [15,16] once the electron relaxation time is known. According to the BoltzTrap2 code [49], the transport electron density of states are defined as

$$\Phi(\varepsilon) \;= \frac{1}{3} tr \left[ \int \sum_i v_i^{\alpha}(\boldsymbol{k}) v_i^{\beta}(\boldsymbol{k}) \, \delta(\varepsilon - \varepsilon_i(\boldsymbol{k})) \frac{d\boldsymbol{k}}{8\pi^3} \right] \qquad \textbf{Eq. 19}$$

where mathematical operator $tr$ means to find the trace of a tensor. The calculated transport electron density of states for the two phases of $Ca_3Ru_2O_7$ are illustrated in **Figure 4**a. Compared with the plot of the DOS's given in Figure 3a, the pseudogap behavior [5,8] is more evident in the plot of transport electron density of states, i.e. a deep dip structure for the AFM-*b* phase vs a shallow structure for the AFM-*a* phase at the Fermi energy, attributed to the significant differences of the electron group velocities between the two phases.

With the transport electron density of states and the electron relaxation time in hand, we can now investigate the electrical conductive properties and understand the *T*-dependent gapping [5]. According to the Cutler-Mott theory [58], the electrical conductivity is formulated as

$$\sigma = \int_{-\infty}^{\infty} \sigma'(\varepsilon) d\varepsilon \qquad \textbf{Eq. 20}$$

where $\sigma'(\varepsilon)$ is a kinetic coefficient called the energy-dependent differential electrical conductivity which is related to the transport density of states in **Eq. 19** by

$$\sigma'(\varepsilon) = \frac{e^2}{k_B T V} f(1 - f) \Phi(\varepsilon) \tau \qquad \textbf{Eq. 21}$$

Again, it is observed that only the electronic states with energies around the Fermi level can contribute the electrical conductive properties dictated by the factor of $f(1-f)$. In **Figure 4**b-e, we choose $T$=10, 50, 90, and 300 K to demonstrate the evolutions of the calculated $\sigma'(\varepsilon)$ for the two phases of $Ca_3Ru_2O_7$.

Finally, the calculated electrical resistivities (the inverse of the conductivity given in **Eq. 20**) of the AFM-$b$ and AFM-$a$ phases for $Ca_3Ru_2O_7$ are compared with experiment [8] in **Figure 4**f. Note that a fair comparison with experiment should be only made up to the Néel temperature of 56 K. By experiment [8,53], above 56 K $Ca_3Ru_2O_7$ is paramagnetic which is not handled in the present work.

## 7    Summary

First-principles calculations based on density functional theory are carried out for the AFM-$b$ and AFM-$a$ phases of $Ca_3Ru_2O_7$.   For the thermodynamic properties at finite temperature, the lattice vibration was handled by phonon approach, and the thermal electron excitation was treated by Mermin's finite temperature DFT approach. For the electron transport properties, the Boltzmann transport equation was solved using the BoltzTraP2 code. The calculated heat capacities agree well with experimental data. Furthermore, a model for estimating the electron relaxation time was proposed so that one can estimate the temperature dependence of the electrical conductivity. The approach has been implemented in the BoltzTraP2 code. Application of the model to the AFM-$b$ and AFM-$a$ phases of $Ca_3Ru_2O_7$ gives rise to promising results when compared with experiment for the temperature dependences of the electrical conductivity.

**Acknowledgements**

This work was supported from the Computational Materials Sciences Program funded by the US Department of Energy, Office of Science, Basic Energy Sciences, under Award Number DE-SC0020145 (Wang, Xiong, Yang, Gopalan, Dabo, and Chen). YW is also partially supported by the Hamer Foundation through the Hamer Professorship. Partially support was from National Science Foundation (NSF) through Grant No. CMMI-1825538 (Wang and Liu). First-principles calculations were carried out partially on the LION clusters at the Pennsylvania State University supported by the Materials Simulation Center and the Research Computing and Cyberinfrastructure unit at the Pennsylvania State University, partially on the resources of NERSC supported by the Office of Science of the US Department of Energy under contract No. DE-AC02-05CH11231, and partially on the resources of Extreme Science and Engineering Discovery Environment (XSEDE) supported by NSF with Grant No. ACI-1053575. We would like to thank Prof. Xianglin Ke for providing his calorimetric heat capacity data.

# References

[1] Y. Yoshida, I. Nagai, S.-I. Ikeda, N. Shirakawa, M. Kosaka, and N. Môri, *Quasi-Two-Dimensional Metallic Ground State of Ca 3 Ru 2 O 7*, Phys. Rev. B **69**, 220411 (2004).

[2] W. X. Zhou and A. Ariando, *Review on Ferroelectric/Polar Metals*, Jpn. J. Appl. Phys. **59**, SI0802 (2020).

[3] Y. Yoshida, S. I. Ikeda, N. Shirakawa, M. Hedo, and Y. Uwatoko, *Magnetic Properties of Ca3Ru2O7 under Uniaxial Pressures*, J. Phys. Soc. Japan **77**, 093702 (2008).

[4] I. Marković et al., *Electronically Driven Spin-Reorientation Transition of the Correlated Polar Metal Ca3Ru2O7*, Proc. Natl. Acad. Sci. U. S. A. **117**, 15524 (2020).

[5] Y. Yuan, P. Kissin, D. Puggioni, K. Cremin, S. Lei, Y. Wang, Z. Mao, J. M. Rondinelli, R. D. Averitt, and V. Gopalan, *Ultrafast Quasiparticle Dynamics in the Correlated Semimetal ${\rm Ca}_3 {\rm Ru}_2 {\rm O}_7$*, Phys. Rev. B **99**, 155111 (2019).

[6] D. Puggioni, M. Horio, J. Chang, and J. M. Rondinelli, *Cooperative Interactions Govern the Fermiology of the Polar Metal Ca3Ru2O7*, Phys. Rev. Res. **2**, 023141 (2020).

[7] X. N. Lin, Z. X. Zhou, V. Durairaj, P. Schlottmann, and G. Cao, *Colossal Magnetoresistance by Avoiding a Ferromagnetic State in the Mott System Ca3Ru2O7*, Phys. Rev. Lett. **95**, (2005).

[8] J. S. Lee, S. J. Moon, B. J. Yang, J. Yu, U. Schade, Y. Yoshida, S. I. Ikeda, and T. W. Noh, *Pseudogap Dependence of the Optical Conductivity Spectra of Ca3Ru2O7: A Possible Contribution of the Orbital Flip Excitation*, Phys. Rev. Lett. **98**, 097403 (2007).

[9] P. Hohenberg and W. Kohn, *Inhomogeneous Electron Gas*, Phys. Rev. **136**, B864 (1964).

[10] W. Kohn and L. J. Sham, *SELF-CONSISTENT EQUATIONS INCLUDING EXCHANGE AND CORRELATION EFFECTS*, Phys. Rev. **140**, A1133 (1965).

[11] Y. Wang, Z.-K. Liu, and L.-Q. Chen, *Thermodynamic Properties of Al, Ni, NiAl, and Ni3Al from First-Principles Calculations*, Acta Mater. **52**, 2665 (2004).

[12] Z. K. Liu and Y. Wang, *Computational Thermodynamics of Materials* (Cambridge University Press, Cambridge, UK, 2016).

[13] Y. Wang, L.-Q. Chen, and Z.-K. Liu, *YPHON: A Package for Calculating Phonons of Polar Materials*, Comput. Phys. Commun. **185**, 2950 (2014).

[14] Y. Wang, S. L. Shang, H. Zhang, L. Q. Chen, and Z. K. Liu, *Thermodynamic Fluctuations in Magnetic States: Fe3Pt as a Prototype*, Philos. Mag. Lett. **90**, 851 (2010).

[15] T. J. Scheidemantel, C. Ambrosch-Draxl, T. Thonhauser, J. V Badding, and J. O. Sofo, *Transport Coefficients from First-Principles Calculations*, Phys. Rev. B **68**, 125210 (2003).

[16] G. D. Mahan and J. O. Sofo, *The Best Thermoelectric*, Proc. Natl. Acad. Sci. **93**, 7436 (1996).

[17] J. Bardeen and W. Shockley, *Deformation Potentials and Mobilities in Non-Polar Crystals*, Phys. Rev. **80**, 72 (1950).

[18] A. J. Hong, L. Li, R. He, J. J. Gong, Z. B. Yan, K. F. Wang, J. M. Liu, and Z. F. Ren, *Full-Scale Computation for All the Thermoelectric Property Parameters of Half-Heusler Compounds*, Sci. Rep. **6**, 1 (2016).

[19] M. Mukherjee, S. Satsangi, and A. K. Singh, *A Statistical Approach for the Rapid Prediction of Electron Relaxation Time Using Elemental Representatives*, Chem. Mater.

**32**, 6507 (2020).

[20]   R. Farris, M. B. Maccioni, A. Filippetti, and V. Fiorentini, *Theory of Thermoelectricity in Mg 3 Sb 2 with an Energy- and Temperature-Dependent Relaxation Time*, J. Phys. Condens. Matter **31**, 065702 (2019).

[21]   Y. Katsura et al., *Data-Driven Analysis of Electron Relaxation Times in PbTe-Type Thermoelectric Materials*, Sci. Technol. Adv. Mater. **20**, 511 (2019).

[22]   S. Poncé, W. Li, S. Reichardt, and F. Giustino, *First-Principles Calculations of Charge Carrier Mobility and Conductivity in Bulk Semiconductors and Two-Dimensional Materials*, Reports on Progress in Physics.

[23]   Y. Wang, Y.-J. Hu, B. Bocklund, S.-L. Shang, B.-C. Zhou, Z.-K. Liu, and L.-Q. Chen, *First-Principles Thermodynamic Theory of Seebeck Coefficients*, Phys. Rev. B **98**, 224101 (2018).

[24]   Y. Wang, J. P. S. Palma, S.-L. Shang, L.-Q. Chen, and Z.-K. Liu, *Lorenz Number and Electronic Thermoelectric Figure of Merit: Thermodynamics and Direct DFT Calculations*, http://arxiv.org/abs/2010.00664 (2020).

[25]   Y. Wang, J. J. Wang, H. Zhang, V. R. Manga, S. L. Shang, L.-Q. Chen, and Z.-K. Liu, *A First-Principles Approach to Finite Temperature Elastic Constants*, J. Phys. Condens. Matter **22**, 225404 (2010).

[26]   Y. Wang, R. Ahuja, and B. Johansson, *Mean-Field Potential Approach to the Quasiharmonic Theory of Solids*, Int. J. Quantum Chem. **96**, 501 (2004).

[27]   N. D. Mermin, *Thermal Properties of the Inhomogeneous Electron Gas*, Phys. Rev. **137**, A1441 (1965).

28    T. Jarlborg, E. G. Moroni, and G. Grimvall, *α-γ Transition in Ce from Temperature-Dependent Band-Structure Calculations*, Phys. Rev. B **55**, 1288 (1997).

29    A. Sommerfeld, *Thermodynamics and Statistical Mechanics* (Academic Press, 1964).

30    A. K. McMahan and M. Ross, *High-Temperature Electron-Band Calculations*, Phys. Rev. B **15**, 718 (1977).

31    V. Fiorentini, R. Farris, E. Argiolas, and M. B. MacCioni, *High Thermoelectric Figure of Merit and Thermopower in Layered Perovskite Oxides*, Phys. Rev. Mater. **3**, 022401 (2019).

32    B. K. Ridley, *Polar-Optical-Phonon and Electron-Electron Scattering in Large-Bandgap Semiconductors*, J. Phys. Condens. Matter **10**, 6717 (1998).

33    L. Lorenz, *Determination of the Degree of Heat in Absolute Measure*, Ann. Phys. Chem. **147**, 429 (1872).

34    P. J. Price, *The Lorenz Number*, IBM J. Res. Dev. **1**, 147 (1957).

35    G. S. Kumar, G. Prasad, and R. O. Pohl, *Review Experimental Determinations of the Lorenz Number*, J. Mater. Sci. **28**, 4261 (1993).

36    M. Thesberg, H. Kosina, and N. Neophytou, *On the Lorenz Number of Multiband Materials*, Phys. Rev. B **95**, 125206 (2017).

37    H. S. Kim, Z. M. Gibbs, Y. Tang, H. Wang, and G. J. Snyder, *Characterization of Lorenz Number with Seebeck Coefficient Measurement*, APL Mater. **3**, 041506 (2015).

38    X. Wang, V. Askarpour, J. Maassen, and M. Lundstrom, *On the Calculation of Lorenz Numbers for Complex Thermoelectric Materials*, J. Appl. Phys. **123**, 055104 (2018).

[39]    K. P. Ong, D. J. Singh, and P. Wu, *Analysis of the Thermoelectric Properties of N-Type ZnO*, Phys. Rev. B - Condens. Matter Mater. Phys. **83**, 115110 (2011).

[40]    A. Bulusu and D. G. Walker, *Review of Electronic Transport Models for Thermoelectric Materials*, Superlattices Microstruct. **44**, 1 (2008).

[41]    A. H. Wilson, *Theory of Metals* (Cambridge University Press, Cambridge, 1953).

[42]    J. Mao, Z. Liu, J. Zhou, H. Zhu, Q. Zhang, G. Chen, and Z. Ren, *Advances in Thermoelectrics*, Adv. Phys. **67**, 69 (2018).

[43]    J. Yang, R. Liu, Z. Chen, L. Xi, J. Yang, W. Zhang, and L. Chen, *Power Factor Enhancement in Light Valence Band P-Type Skutterudites*, Appl. Phys. Lett. **101**, 022101 (2012).

[44]    J. M. Ziman, *Electrons and Phonons: The Theory of Transport Phenomena in Solids* (Clarendon Press, 2001).

[45]    G. Kresse and J. Furthmüller, *Efficiency of Ab-Initio Total Energy Calculations for Metals and Semiconductors Using a Plane-Wave Basis Set*, Comput. Mater. Sci. **6**, 15 (1996).

[46]    G. Kresse and D. Joubert, *From Ultrasoft Pseudopotentials to the Projector Augmented-Wave Method*, Phys. Rev. B **59**, 1758 (1999).

[47]    J. P. Perdew and A. Zunger, *Self-Interaction Correction to Density-Functional Approximations for Many-Electron Systems*, Phys. Rev. B **23**, 5048 (1981).

[48]    S. L. Dudarev, G. A. Botton, S. Y. Savrasov, C. J. Humphreys, and A. P. Sutton, *Electron-Energy-Loss Spectra and the Structural Stability of Nickel Oxide: An LSDA+U Study*, Phys. Rev. B **57**, 1505 (1998).

49    G. K. H. Madsen, J. Carrete, and M. J. Verstraete, *BoltzTraP2, a Program for Interpolating Band Structures and Calculating Semi-Classical Transport Coefficients*, Comput. Phys. Commun. **231**, 140 (2018).

50    P. Virtanen, R. Gommers, T. E. Oliphant, M. Haberland, T. Reddy, D. Cournapeau, E. Burovski, P. Peterson, W. Weckesser, and J. Bright, *SciPy 1.0: Fundamental Algorithms for Scientific Computing in Python*, Nat. Methods **17**, 261 (2020).

51    Y. Wang, M. Liao, B. J. Bocklund, P. Gao, S. L. Shang, H. Kim, A. M. Beese, L. Q. Chen, and Z. K. Liu, *DFTTK: Density Functional Theory ToolKit for High-Throughput Lattice Dynamics Calculations*, Calphad **75**, 102355 (2021).

52    S.-L. Shang, Y. Wang, D. Kim, and Z.-K. Liu, *First-Principles Thermodynamics from Phonon and Debye Model: Application to Ni and Ni3Al*, Comput. Mater. Sci. **47**, 1040 (2010).

53    X. Ke, J. Peng, W. Tian, T. Hong, M. Zhu, and Z. Q. Mao, *Commensurate-Incommensurate Magnetic Phase Transition in the Fe-Doped Bilayer Ruthenate Ca3Ru2 O7*, Phys. Rev. B - Condens. Matter Mater. Phys. **89**, 220407 (2014).

54    S. Mccall, G. Cao, and J. E. Crow, *Impact of Magnetic Fields on Anisotropy in Ca 3 Ru 2 O 7*, APS **67**, (2003).

55    C. Kittel, *Introduction to Solid State Physics* (Wiley, 2005).

56    G. Cao, S. McCall, J. E. Crow, and R. P. Guertin, *Observation of a Metallic Antiferromagnetic Phase and Metal to Nonmetal Transition in Ca3Ru2O7*, Phys. Rev. Lett. **78**, 1751 (1997).

57    Z.-K. Liu and Y. Wang, *Computational Thermodynamics of Materials* (Cambridge

University Press, Cambridge, UK, 2016).

[58]     M. Cutler and N. F. Mott, *Observation of Anderson Localization in an Electron Gas*, Phys. Rev. **181**, 1336 (1969).

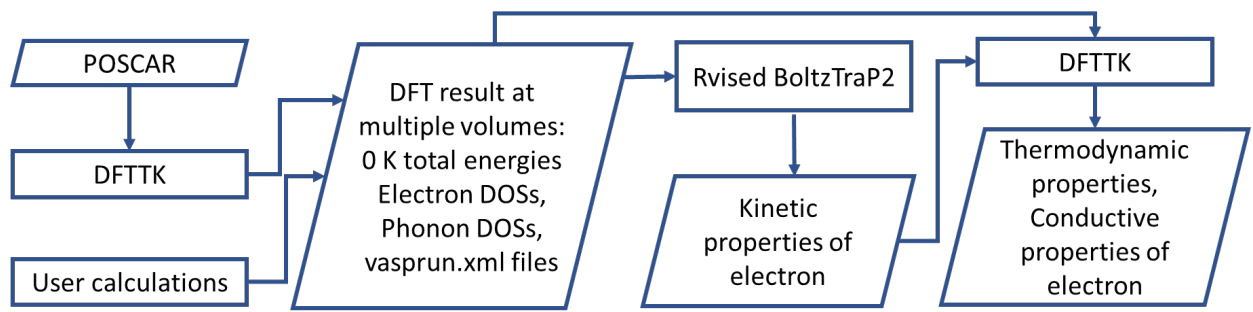

**Figure 1.** Computerization of workflow

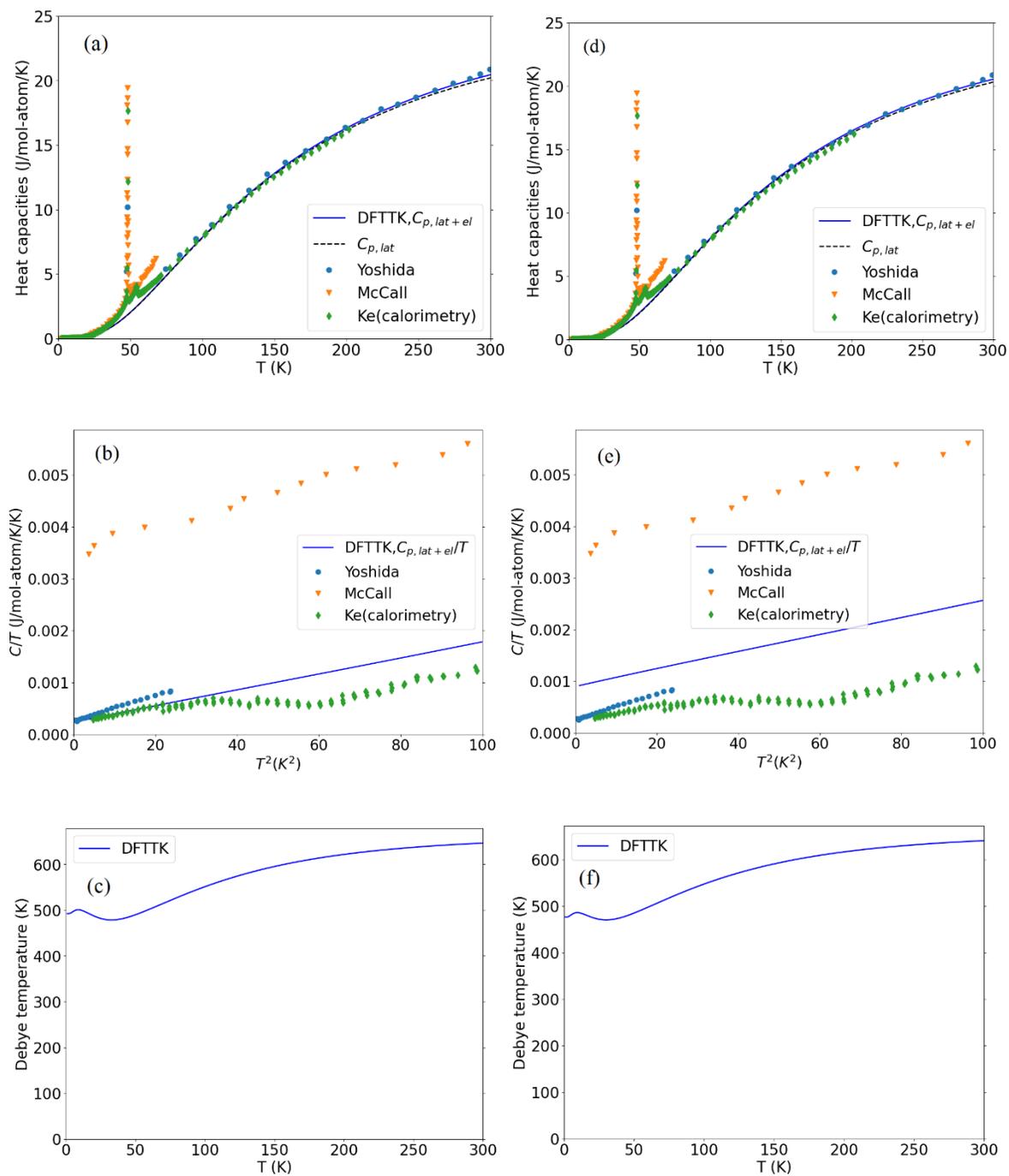

**Figure 2.** Heat capacities, electronic heat capacity coefficients, and Debye temperatures for the AFM-*b* [(a), (b), and (c)] and AFM-*a* [(d), (e), and (f)] phases of Ca$_3$Ru$_2$O$_7$, respectively. The dots

are experimental data [1,53,54]. The dashed lines in the heat capacity plot are for the calculated values without considering the thermal electronic contributions. $C/T$ vs $T^2$ plot represents the analysis of the heat capacity at low temperature heat capacity .

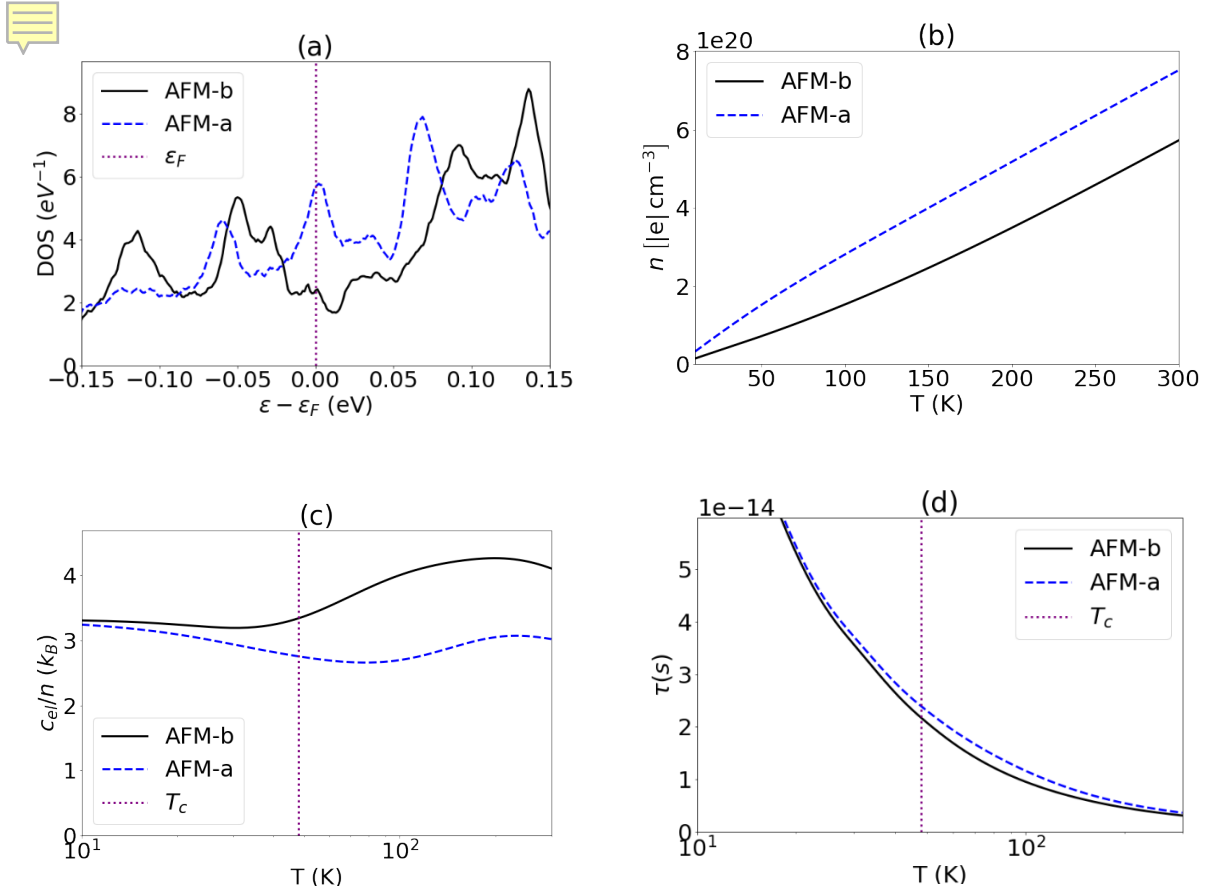

Figure 3. Calculated electronic properties based on the electron density of states for the **AFM-$b$** and **AFM-$a$** phases of Ca$_3$Ru$_2$O$_7$. (a) the electron density of states; (b) the density of the active electronic thermal carriers; (c) $c_{el}/n$ the electronic heat capacity per active electronic thermal carriers; and (d) the relaxation time estimated using Eq. 17 based on $c_{el}/n$ in Eq. 13.

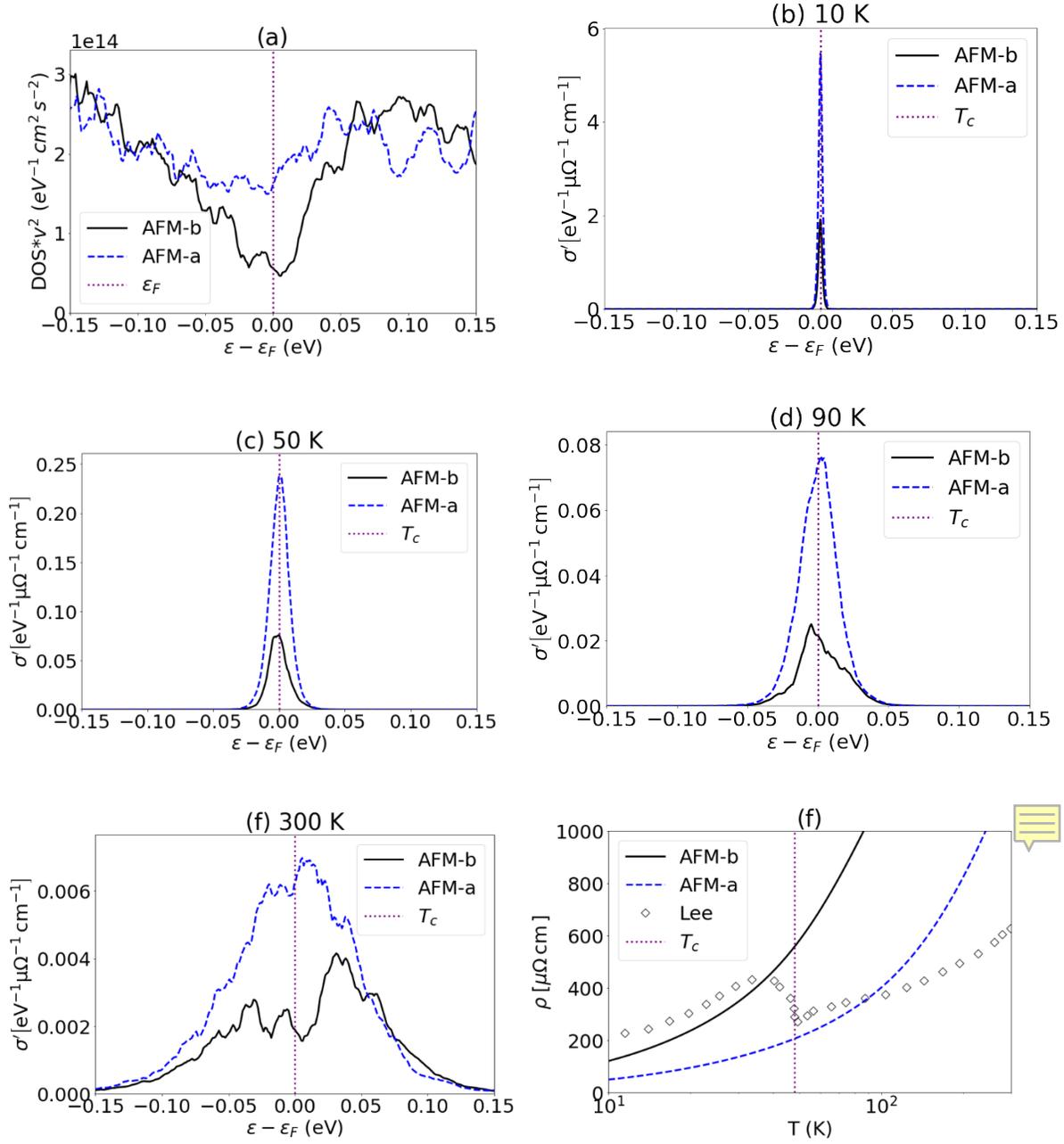

Figure 4. The calculated conductive properties for the **AFM-b** (solid lines) and **AFM-a** (dashed lines) phases of Ca$_3$Ru$_2$O$_7$. (a) the transport density of states of electron as defined in Eq. 20; (b)-(e) the Mott the energy-dependent differential electrical conductivity at 10, 50, 90, and 300 K,

respectively; and (f) the electrical resistivity. The diamonds in (f) are the experimental dc electrical resistivity reported by Lee et al. [8].